# Small and simple systems that favor the arrow of time

Ralph V. Chamberlin, Department of Physics, Arizona State University, Tempe AZ 85287-1504 USA

**Abstract**

The 2nd law of thermodynamics yields an irreversible increase in entropy until thermal equilibrium is achieved. This irreversible increase is often assumed to require large and complex systems to emerge from the reversible microscopic laws of physics. We test this assumption using simulations and theory of a 1D ring of $N$ Ising spins coupled to an explicit heat bath of $N$ Einstein oscillators. The exact entropy of the spins and bath can be calculated for any $N$, with dynamics that is readily altered from reversible to irreversible. We find thermal equilibrium behavior in the thermodynamic limit, and in systems as small as $N = 2$, but only if the microscopic dynamics is intrinsically irreversible.

The second law of thermodynamics is the fundamental barrier that prevents us from going back in time. Indeed, this "2nd law" causes the unidirectional increase in entropy needed for the "arrow of time" [1-6] and the maximum in entropy for thermal equilibrium. Because other basic laws of physics are reversible, the 2nd law dominates our daily lives. However, most (but not all [7-9]) scientists believe that irreversible behavior emerges from reversible microscopic dynamics only for systems that are large and complex. As an early example, Boltzmann often sought to obtain the irreversible 2nd law from reversible Newtonian kinetics, with mixed success. His goal was to justify the maximum in entropy needed for thermal equilibrium, now embodied by Boltzmann's factor $p_i \propto e^{-E_i/kT}$, where Boltzmann's constant ($k$) connects internal energies ($E_i$) to an ideal heat bath at temperature $T$. (An ideal heat bath is effectively infinite, with weak but essentially instantaneous thermal coupling to the system [10].) Although experiments must ultimately decide the range of 2nd-law behavior [11-14], the models we describe here have the advantage of dynamics that is easily changed from reversible to irreversible, system sizes that are precisely fixed, and entropy of the entire system (including heat bath) that can be calculated exactly at every step.

Computer simulations provide unique insight into how macroscopic behavior emerges from microscopic dynamics [15-17]. In fact, the first molecular dynamics (MD) simulations on a digital computer surprised Enrico Fermi and his team by showing that small anharmonic systems governed by Newton's laws fail to reach thermal equilibrium [18]. Although MD simulations of larger systems usually exhibit thermal-equilibrium averages [19,20], when Fermi's ideas are extended to study dynamics, many systems have energy fluctuations that diverge from standard statistical mechanics as $T \to 0$ [21]. This disconnect between Boltzmann's factor and Newton's laws arises when conservation of local energy overwhelms the weak coupling to the heat bath. Here we study simpler systems as a function of size and coupling to the bath. We find that an intrinsically irreversible step in the microscopic dynamics is needed for the 2nd law.

All systems studied here are based on the Ising model, consisting of interacting binary degrees of freedom ("spins") on a periodic lattice [22]. We start with the one-dimensional (1D) Ising model having $N$ spins in a 1D "ring" (to avoid anomalous endpoints). The potential energy is $U = -J\sum_{i=1}^{N} b_i\sigma_i\sigma_{i+1}$. Here, $J$ is an interaction energy, and $b_i$ governs whether this interaction ("bond") occurs between neighboring spins at lattice sites $i$ and $i+1$, with $b_N$ between spins at $i = N$ and $i = 1$. Thus, bonded spins have interaction energy $-J$ if they are aligned or $+J$ if they are anti-aligned, or 0 if they are unbonded. In the standard Ising model, all spins are bonded to all nearest neighbors, $b_i \equiv 1$. Here we usually assume that broken bonds ($b_i = 0$) occur intermittently, allowing a thermal-equilibrium distribution of bonds $B = \sum_{i=1}^{N} b_i < N$. These broken bonds are needed to maximize entropy for the 2nd law, plus they add complexity via a third energy value between neighboring spins.

The 1D Ising model can be treated analytically for systems of any size using Boltzmann's entropy [23]

$$S_U/k = \ln\left[\frac{N!2^{n_0}}{(N-n_0-n_x)!n_0!n_x!}\right]. \quad \text{Eq. (1)}$$

Here $n_x$ is the number of high-energy bonds between anti-aligned spins and $n_0$ is the number of non-interacting bonds, leaving $N - n_0 - n_x$ low-energy bonds between aligned spins. Note that the term $2^{n_0}$ is a consequence of $n_0$ separate segments of spins (having non-bonded endpoints) in a continuous ring, which is replaced by $2^{n_0+1}$ for free boundaries, or when $n_0 = 0$. However, the special case of $n_0 = 0$ is negligible if $N \gg 10$ and $T$ is not too low. Then, the system rarely fluctuates far from the condition that maximizes the entropy of mixing, $n_0/N = 1/2$. At the energies investigated here, equilibrium values are: $n_0/N = 0.51 - 0.57$. In terms of the bond types given in Eq. (1), the potential energy becomes:

$$U/J = -(N - n_0 - 2n_x) \quad \text{Eq. (2)}$$

Although simulations of the Ising model usually utilize algorithms based on Boltzmann's factor, we argue that such simulations yield both a conceptual and physical mismatch by assuming that an effectively-infinite heat bath instantly couples to a small set of spins. A more balanced approach, introduced by Creutz [24], couples the spins to a similarly-sized set of "demons" (Einstein oscillators, all having the same spacing between discrete energy levels). In Creutz's model, each demon serves as a local heat bath that exchanges energy with nearby spins; thus, the demons serve as a source of kinetic energy, $K$. This model enhances the realism of simple simulations by requiring that total energy ($E = U + K$) is always explicitly conserved, and by ensuring that the spin system and its heat bath have similar sizes and impact on each other. Of special interest is when the Creutz model is made reversible [25], where simulations can be run for an unlimited sequence of steps, then if reversed, after the same number of steps the system returns to its exact initial state. Such precise reversibility is rare in simulations of complex systems, where round-off error and sensitivity to initial conditions often yield divergent trajectories [26,27]. Although many types of cellular automata show reversibility [28], the Creutz model also has discrete states for its heat bath, yielding thermal statistics for small systems and thermodynamic behavior for large systems.

Unlike most simulations of Ising spins that utilize a fixed temperature from an ideal heat bath, simulations of the Creutz model conserve energy, so that $T$ is found from thermal averages. Consider $N$ sources of $K$, each with energy-level spacing $J$ to match the spins. If each $K$ was coupled to an ideal heat bath, its average energy would be given by Bose-Einstein statistics: $K_\infty = J/(e^{J/kT} - 1)$. Inverting this equation yields the average temperature: $kT/J = 1/\ln(1 + J/K_\infty)$. Although finite-size effects in small systems corrode the concept of $T$, the exact Boltzmann's entropy for $N$ sources of $K$ can be written for systems of any size [29]:

$$S_K/k = \ln\left[\frac{(K+N-1)!}{K!(N-1)!}\right] \qquad \text{Eq. (3)}$$

In fact, we find accurate thermal averages for $N \geq 2$ from counting the ways that $E = U + K$ can be distributed between $U$ and $K$. Specifically, the entropies, Eq. (1) and Eq. (3), yield the multiplicities, $W_U = e^{S_U/k}$ and $W_K = e^{S_K/k}$, and the net probabilities:

$$p_{U,K} = (W_U * W_K)/\sum_{U,K}(W_U * W_K). \qquad \text{Eq. (4)}$$

Table 1 gives key properties of the $N = 2$ system having total energy $E/NJ \equiv 1$. Even in a relatively short simulation, all possible spin states and bonds are likely to occur, with weighted averages given in the bottom row of Table 1. Similar tables can be constructed for systems having other sizes and energies, but such tables become unwieldy for large systems.

We simulated Creutz-like models as detailed in the Supplemental Material. The main part of Fig. 1 shows the resulting total entropy density as a function of time for various simulations of a system having $N = 2x10^6$ and $E/NJ \equiv 1$. Figures 1 A, B, and C show, respectively, the dynamics at the beginning, near the middle (vertical scale $x10^5$), and end of the simulations. The upper three sets of data in Figs. 1 A and C also have an expanded vertical scale ($x10^2$), with a common offset in both cases. Two sets of simulations are reversible, with the dynamics of each step governed by either the local $K$ (green and grey), or global $K$ (blue and cyan) chosen from the entire bath using randomly-ordered (but fixed) arrays. The third set of simulations (red and magenta) are irreversible, with global $K$ chosen using a new random number each step. Main colors (green, blue, and red) come from the initial simulation, with spins always starting in the same highly-aligned state. Secondary colors (grey, cyan, and magenta) come from averaging three subsequent simulations of each type, alternating reversible then irreversible dynamics, with spins starting in the final state of the previous simulation. All reversible simulations (green, grey, blue, and cyan) start and end in the same state, while irreversible simulations (red and magenta) show no tendency to return to their initial state. Horizontal lines in Fig. 1 B show that when averaged over all intermediate steps, net entropy increases slightly with increasing randomness for the initial simulations (solid), while subsequent simulations (dashed) show a sharp increase from dynamics that is reversible (grey and cyan), to irreversible (magenta). Black lines in Figs. 1 A and C come from the initial simulation of a similar system, with global $K$ and reversible dynamics, but no broken bonds. The peak entropy densities of all other simulations are about $k\ln(2)$ higher, as expected from the entropy of mixing when broken bonds are allowed. Such large increases in entropy indicate that real systems must also intermittently break their interactions, if physically feasible.

Four lines in Fig. 1 A (black, green, blue, and red) show that the entropy during each initial simulation rises sharply from an initial state, with rates that increase with increasing randomness. Specifically, for the entropy to reach $95\%$ of the maximum value, irreversible dynamics (red) requires a single sweep, while reversible dynamics requires seven sweeps with global $K$ (blue) and eleven sweeps with local $K$ (green); whereas systems with no broken bonds (black) never approach this maximum. Two lines in Figs. 1 A and C (red and magenta) show that systems with intrinsic randomness are irreversible. All other lines are reversible, precisely following every step back in time, even after more than $2.62x10^{11}$ steps.

Figure 1 B displays these same simulations over an intermediate interval of times on a greatly expanded scale. Although initial simulations (green, blue, and red) have considerable overlap in their fluctuations, their time-averaged entropies (solid horizontal lines) increase significantly with increasing randomness (standard errors are less than the line thickness). Furthermore, subsequent simulations exhibit a large jump in entropy from dynamics that is reversible (grey and cyan) to irreversible (magenta), which is even clearer when time-averaged (dashed horizontal lines). Thus, Fig. 1 establishes that the total entropy from irreversible dynamics is significantly and persistently higher than that from reversible dynamics. Moreover, if the spins are initially in an equilibrium state due to irreversible dynamics, this entropy evolves to a lower value if the dynamics becomes reversible.

The inset in Fig. 1 shows power spectral density (*PSD*) as a function of relative frequency ($f$) using the same sets of simulations and line colors given in the main part of the figure. Each *PSD* is found by taking the absolute value squared of the Fourier transform of $\sqrt{K}$. The two sets of irreversible simulations (magenta) often overlap, giving a measure of the uncertainty. These *PSD* decrease monotonically with increasing *f* for $10\log(f) > 30$, consistent with the overdamped relaxation shown in Fig. 1 A. Lines from reversible simulations having local $K$ (grey) and global $K$ (cyan) overlap at high frequencies. Both of these *PSD* increase monotonically with increasing *f* for $10\log(f) > 30$, reaching a peak at maximum *f*, consistent with the fast oscillations shown in Figs. 1 A and C. With decreasing *f*, reversible dynamics utilizing local $K$ (grey) shows a 1/*f*-like divergence, indicative of slow energy diffusion at long times.

Figure 2 A shows moving averages of the total entropy from the simulations of Fig. 1, with Figs. 2 B and C from the separate entropies of $K$ and $U$, respectively. Each symbol comes from averaging $10^4$ sweeps, positioned at their median time. Global $K$ is used for both reversible dynamics (black squares) and irreversible dynamics (red circles). Open symbols come from initial simulations, with closed symbols from averaging three pairs of subsequent simulations. Error bars (visible when larger than the symbol size) give the standard error from averaging the subsequent simulations. Total entropies (Fig. 2 A) of the initial reversible simulation (open squares) are nearly as high as the initial irreversible simulation (open

circles), consistent with the same simulations in Fig. 1 B (blue and red lines, respectively). Entropies of subsequent reversible simulations in Fig. 2 A (filled squares) are sharply lower than irreversible simulations (filled circles), again consistent with the behavior in Fig. 1 B (cyan and magenta lines, respectively). The total entropy in Fig. 2 A tracks the entropy of $K$ in Fig. 2 B, but mirrors the entropy of $U$ in Fig. 2 C. Thus, reversible dynamics fails to reach maximum entropy because of the explicit heat bath, which is absent in simulations utilizing an ideal heat bath where total entropy is not calculated. Our results are consistent with MD simulations showing deviations from standard statistical mechanics [21], attributable to the explicit conservation of energy and intrinsic reversibility of Newton's laws.

Now focus on the behavior during the middle third of Fig. 2. During these middle times, all simulations have the average rate of bond-change attempts reduced to $1/10^{th}$ the rate for spin-change attempts. Note that the total entropy is significantly altered only for reversible dynamics. Interestingly, this total entropy is reduced during the initial simulation, but increased during subsequent simulations. Thus, reversible dynamics yields non-equilibrium steady states with entropy that depends on the relative time scales of the dynamics, and on the initial conditions [30,31]. Whereas, entropy is significantly and consistently higher for irreversible dynamics.

Symbols in Fig. 2 (D) show time dependences of the ratio of probabilities from adjacent levels in $K$. Specifically: $\ln(p_i/p_{i+1})$ with $i = 0$ (squares), $i = 1$ (circles), $i = 2$ (up triangles), and $i = 3$ (down triangles). Thus, if Boltzmann's factor can be used for $K$, each symbol gives $\Delta K/kT$, where $\Delta K = K_{i+1} - K_i = J$. Coinciding red symbols indicate that a single $T$ applies to all levels only for irreversible dynamics of large systems. Black symbols show that reversible dynamics requires one value of effective $T$ for $i = 0$ and $i = 2$, another value for $i = 1$ and $i = 3$, with additional values for middle times when bond-change rates are reduced. The concept of a single $T$ also fails for irreversible dynamics in smaller systems ($N = 128$, green symbols), but such finite-size effects are expected when a fixed total energy is insufficient to thermally occupy higher energy levels. Thus, Boltzmann's factor applies only in the thermodynamic limit, and only if the dynamics is intrinsically irreversible.

Figure 3 displays various entropy densities, and their differences, as a function of $N$. The main plot has logarithmic axes with $2 \leq N \leq 2x10^6$, while the inset has linear axes with $2 \leq N \leq 16$. Line and symbol colors identify the total energy during each simulation (see legends). The uppermost lines (dash-dotted) show that $S_{irr}/Nk$ increases with increasing $N$, approaching a constant value at large $N$. Symbols in the inset show that $S_{irr}/Nk$ from simulations (circles) coincide with $S_{theo}/Nk$ from theory (squares), found from summing the entropies of all states weighted by their multiplicities, similar to those given in Table 1. The inset also shows $(S_{irr} - S_{theo})/S_{theo}$ (triangles) in percent. Although Boltzmann's factor cannot describe such small systems, for intrinsically irreversible dynamics, thermal equilibrium from the entropy-weighted sum over all states remains valid down to $N = 2$. Of course, thermal equilibrium fails for $N = 1$ because there can be no randomness in choosing the single $K$.

Symbols in the main part of Fig. 3 show the size dependence of the difference between entropies, $(S_{irr} - S_{rev})/Nk = \Delta S/Nk$. Symbol color and type identify the total energy and heat bath, given in the legend. Lines, from fits to the data using $\Delta S/Nk = s/N + c$ with $s$ and $c$ constants, show general agreement with the behavior from local (dashed) and global (solid) $K$. Irreversible dynamics increases the entropy, $\Delta S > 0$, except for three blue triangles missing from Fig. 3 where $N = (1\ or\ 4)x10^4$ with $E/NJ \equiv 2$. Having $\Delta S < 0$ can be understood from Fig. 2 D by the number of values of effective $T$ needed to describe the dynamics, which changes from many values for small systems (green symbols), to two values for large reversible systems (black symbols) but only one value for irreversible dynamics (red symbols). More importantly, even as $N \to 2$ where $\Delta S/Nk > 0.01$, the inset shows that only $S_{irr}$ matches calculated multiplicities. Most importantly, when $N > 10^4$, $\Delta S/Nk$ tends to increase with increasing $N$, opposite to the behavior needed for $S_{rev}$ to approach $S_{irr}$ in the thermodynamic limit.

We identify oscillations in local energy as the primary cause of non-thermal behavior during reversible dynamics. Grey and cyan lines in Fig. 1 manifest these oscillations in the entropy at the start (A) and end (C) of the simulations, and in the inset from the peak at highest frequencies. These oscillations occur as fluctuations in local energy are first traded back-and-forth between $U$ and $K$ before dissipating via bonded spins. Similar fast oscillations in energy during MD simulations cause analogous deviations from standard statistical mechanics [21]. Thus, Boltzmann's factor fails to describe fluctuations in large and complex systems governed by Newton's laws, and in our simple systems governed by reversible dynamics. In both cases, the failure of standard statistical mechanics can be attributed to conservation of local energy overwhelming the weak coupling to an explicit heat bath. However, when our simple systems have intrinsically irreversible dynamics, red symbols and lines in Figs. 1 and 2 A show significantly higher entropy, and a well-defined $T$ as $N \to \infty$ (Fig. 2 D). Furthermore, the inset of Fig. 3 shows thermal-equilibrium behavior for systems as small as $N = 2$, but only if the dynamics is intrinsically irreversible.

We speculate that intrinsic irreversibility in real systems may come from wavefunction collapse when coupling to a bath. In our case, each step involves one spin coupling to one $K$, so that $N = 2$ may be related to a double-slit experiment. For reversible dynamics, the choice of $K$ follows a prescribed sequence, similar to always knowing which slit passes the particle. Whereas irreversible dynamics involves intrinsic randomness in the choice of $K$, similar to the intrinsic randomness of wavefunction collapse when a particle is measured as it passes through a slit. Thus, our results suggest that the measurment processes may extend to microscopic systems, such as when a single spin senses a heat bath of at least two particles. Related ideas have previously been proposed for the 2nd law [1,3,8,9].

*Conclusions* – Our simulations reveal some unanticipated results in the thermal behavior of simple systems. An intrinsically irreversible step is needed for maximum entropy in systems of all sizes. Specifically, intrinsically irreversible dynamics is needed for equilibrium thermal statistics from entropy-weighted sums in systems as small as $N = 2$, and for Boltzmann's factor to apply in the thermodynamic limit, $N \gg 10^4$. The failure of standard statistical mechanics to describe reversible dynamics can be attributed to conservation of local energy causing excess fluctuations that are not fully constrained by weak coupling to a realistic heat bath. Because maximum entropy and irreversible behavior are often observed in our daily lives, our results suggest that intrinsic randomness may be common in nature.

I gratefully acknowledge insightful comments from Sumiyoshi Abe, Ruth E. Kastner, Vladimiro Mujica, and George H. Wolf. I also thank Research Computing at Arizona State University for use of their facilities.

***References***

1 Davies, P.C.W. *The Physics of Time Asymmetry*, University of California Press (1974).

2 Lebowitz, J.L. Boltzmann's entropy and time's arrow. *Phys. Today* **46**, 32-38 (1993).

3 Albert, D. Z.; *Time and Chance*, Harvard University Press, Cambridge MA (2000).

4 Schollwöck, U. Why does time have a future? The physical origins of the arrow of time. *Configurations* **23**, 177-195 (2015).

5 't Hooft, G. Time, the arrow of time, and quantum mechanics. *Front. Phys.* **6**, 81 (2018).

6 Roduner, E.; Kruger, T. P. J. The origin of irreversibility and thermalization in thermodynamic processes, *Phys. Rep.* **944**, 1-43 (2022).

7 Wehrl, A. General properties of entropy. *Rev. Mod. Phys.* **50**, 221-260 (1978).

8 Kastner, R. E., On quantum collapse as a basis for the second law of thermodynamics, *Entropy* **19**, 106 (2017)

9 Kastner, R. E., Decoherence and the transactional interpretation, *Int. J. Quant. Found.* **6**, 24-39 (2020)

10 Feynman, R. P.; *Statistical Mechanics*, Perseus Books, Reading MA (1998).

11 Batalhão, T. B.; Souza, A. M.; Sarthour, R. S.; Oliveira, I. S.; Paternostro, M.; Lutz, E.; Serra, R. M. Irreversibility and the arrow of time in a quenched quantum system. *Phys. Rev. Lett.* **115**, 190601 (2015)

12 Zhang, J. W.; Rehan, K.; Li, M.; Li, J. C.; Chen, L.; Su, S.-L.; Yan, L.-L.; Zhou, F.; Feng, M. Single-atom verification of the information-theoretical bound of irreversibility at the quantum level. *Phys. Rev. Res.* **2**, 033082 (2020)

13 Jayaseelan, M.; Manikandan, S. K.; Jordan, A. N.; Bigelow, N. P. Quantum measurement arrow of time and fluctuation relations for measuring spin of ultracold atoms, *Nat. Comm.* **12**, 1847 (2021)

14 Lynn, C. W.; Holmes, C. M.; Bialek, W.; Schwab, D. J. Decomposing the local arrow of time in interacting systems, *Phys. Rev. Lett.* **129**, 118101 (2022)

15 Schroeder, D. V. Interactive molecular dynamics. *Am. J. Phys.* **83**, 210-218 (2015)

16 Hoover, W. G.; Hoover, C. G.; Smith, E. R. Nonequlibrium time reversibility with maps and walks. *Entropy* **24**, 78 (2022)

17 Veszeli, M. T.; Vattay, G. Relaxation of the Ising spin system coupled to a bosonic bath and the time dependent mean field equation. *PLoS One* **17**, e0264412 (2022).

18 Fermi, E.; Pasta, J. R.; Ulam, S. Report LA-1940. Los Alamos Scientific Laboratory (1955).

19 Livi, R.; Pettini, M.; Ruffo, S.; Sparpaglione, M.; Vulpiani, A. Equipartition threshold in nonlinear large Hamiltonian systems: The Fermi-Pasta-Ulam model, *Phys. Rev. A* **31**, 1039-1045 (1985)

20 Matsuyama, H. J.; Konishi, T. Multistage slow relaxation in a Hamiltonian system: The Fermi-Pasta-Ulam model, *Phys. Rev. E* **92**, 022917 (2015).

21 Chamberlin, R.V.; Mujica, V.; Izvekov, S.; Larentzos, J.P. Energy localization and excess fluctuations from long-range interactions in equilibrium molecular dynamics, *Physica* A, **540**, 123228 (2020).

22 Niss, M. History of the Lenz-Ising model 1920-1950: from ferromagnetic to cooperative phenomena, *Arch. Hist. Exact Sci.* **59**, 267-318 (2005) and History of the Lenz-Ising model 1950-1965: from irrelevance to relevance, ibid **63**, 243-287 (2009).

23 Our $S_U/k$ has various differences from $S_c/k$ in Table 3 of Chamberlin, R. V., Clark, M. R., Mujica, V. & Wolf, G. H. Multiscale thermodynamics: Energy, entropy, and symmetry from atoms to bulk behavior. *Symmetry* **13**, 721 (2021). First, $\eta'$ in $S_c/k$ equals $n_0$ here, Second, Table 3 has an error: $(\eta' + 1)\ln(2)$ should be added to $S_c/k$ for the multiplicities of different segments. Finally, because here we use periodic boundary conditions, our $S_U/k$ will have $(n_0)\ln(2)$ added to it if $n_0 > 0$, or $\ln(2)$ added if $n_0 = 0$.

24 Creutz, M. Microcanonical Monte Carlo simulations. *Phys. Rev. Lett.* **50**, 1411-1414 (1983).

25 Creutz, M. Deterministic Ising dynamics. *Ann. Phys.* **167**, 62-72 (1986).

26 Komatsu, N.; Abe, T. Noise-driven numerical irreversibility in molecular dynamics technique. *Comp. Phys. Comm.* **171**, 187-196 (2005).

27 Schroeder, D. V. Interactive molecular dynamics. *Am. J. Phys.* **83**, 210-218 (2015).

28 Kari, J., Theory of cellular automata: A survey, *Theor. Computer Sci*. **334**, 3-33 (2005)

29 Callen, H.B. Thermodynamics and an Introduction to Thermostatistics 2nd ed.; John Wiley & Sons; Hoboken, NJ, USA, (1985).

30 Zhang, F.; Isbister, D. J.; Evans, D. J. Multiple nonequilibrium steady states for one-dimensional heat flow, *Phys. Rev. E* **64**, 021102 (2001).

31 Iwatsuka, T.; Fukai, Y. T.; Takeuchi, K. A. Direct evidence for universal statistic of stationary Kardar-Parisi-Zhang interfaces, *Phys. Rev. Lett.* **124**, 250602 (2020).

| Config. | $n_0$ | $n_x$ | $U/J$ | $S_U/k$ | $K/J$ | $S_K/k$ | $S_E/k$ | $p_{U,K}$ |
|---|---|---|---|---|---|---|---|---|
| ⇑ ● ⇑ ● | 0 | 0 | -2 | ln(2) | 4 | ln(5) | ln(10) | 10/48 |
| ⇑ × ⇓ × | 0 | 2 | 2 | ln(2) | 0 | ln(1) | ln(2) | 2/48 |
| ⇑ ● ⇑ ⨀ | 1 | 0 | -1 | ln(4) | 3 | ln(4) | ln(16) | 16/48 |
| ⇑ × ⇓ ⨀ | 1 | 1 | 1 | ln(4) | 1 | ln(2) | ln(8) | 8/48 |
| ⇑ ⨀ ⇓ ⨀ | 2 | 0 | 0 | ln(4) | 2 | ln(3) | ln(12) | 12/48 |
| | Avg/$N$: | | -0.25 | 0.606504 | 1.25 | 0.593788 | 1.2002914 | |

**TABLE 1.** Properties of $N = 2$ system having total energy $E/NJ \equiv 1$. The first column shows representative configurations for low energy (●), high energy (×), or no energy (⨀) bonds; with spins that may be up (⇑) or down (⇓). The next four columns give the number of no-energy ($n_0$) and high-energy ($n_x$) bonds, $U/J$ (Eq. (2)) and $S_U/k$ (Eq. (1)). The next two columns give $K = E - U$ and $S_K/k$ (Eq. (3)). The final two columns give $S_E = S_U + S_K$ and $p_{U,K}$ (Eq. (4)). The bottom row gives weighted averages.

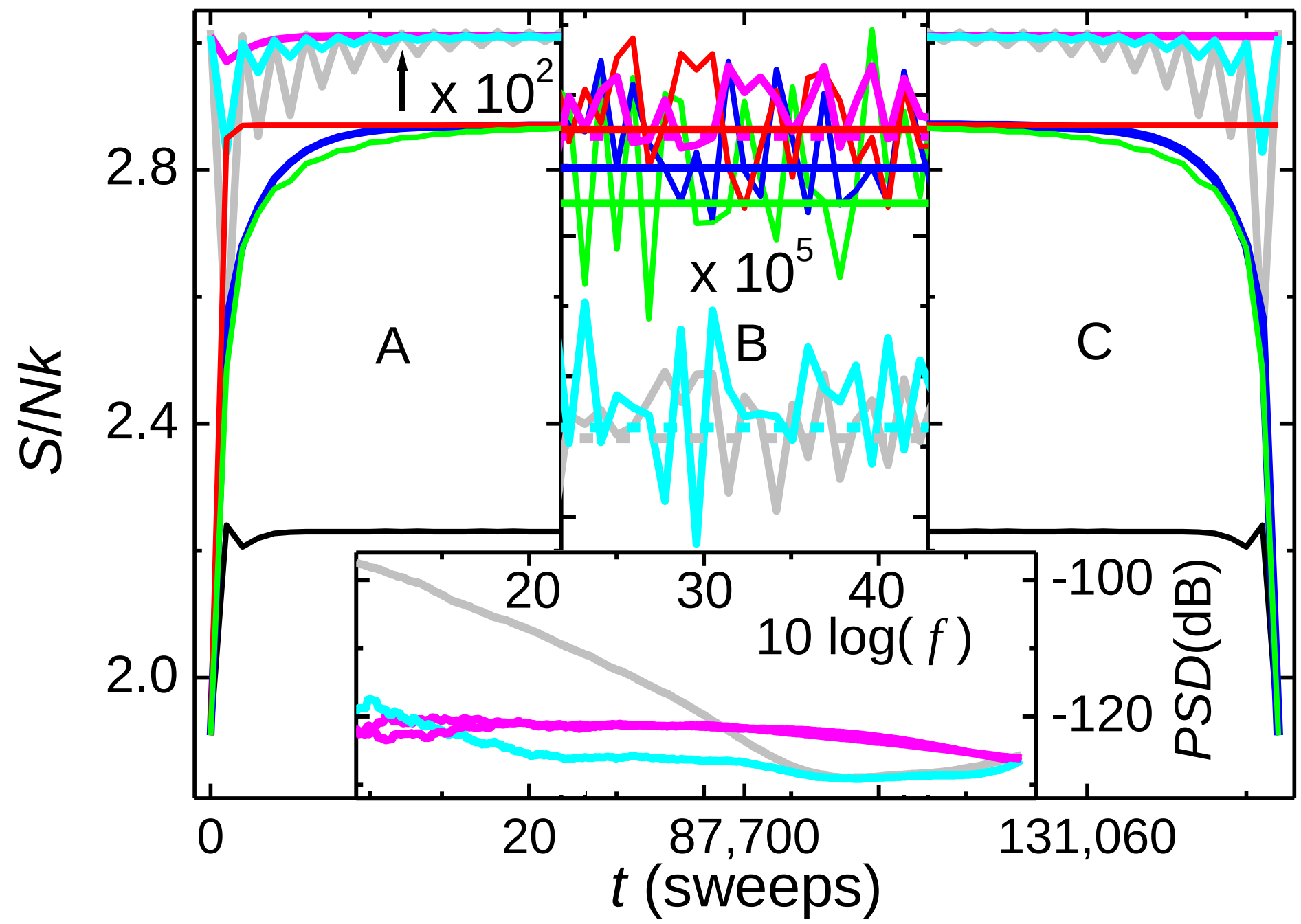

**FIG 1.** Main figures show $S/Nk$ as a function of time at the beginning (A), ending (C), and intermediate (B) times of simulations having $N = 2x10^6$ and $E/NJ \equiv 1$. Line color identifies the type of dynamics and thermal bath for initial simulations (main axis), and when averaged over subsequent simulations on an expanded scale of $x10^2$ in A and C, and $x10^5$ in B, with a common offset in each case. Specifically, red (magenta) lines show initial (averaged) behavior of irreversible simulations. Initial (averaged) reversible dynamics using local $K$ is shown by the green (grey) lines, whereas dynamics using global $K$ is shown by the blue (cyan) lines. Horizontal lines in B show time-averaged values of each type. Black lines in A and C are also from reversible dynamics using global $K$, but with no broken bonds. The inset shows power-spectral densities as a function of relative frequency on a double-logarithmic plot, with line color matching that in the main figures.

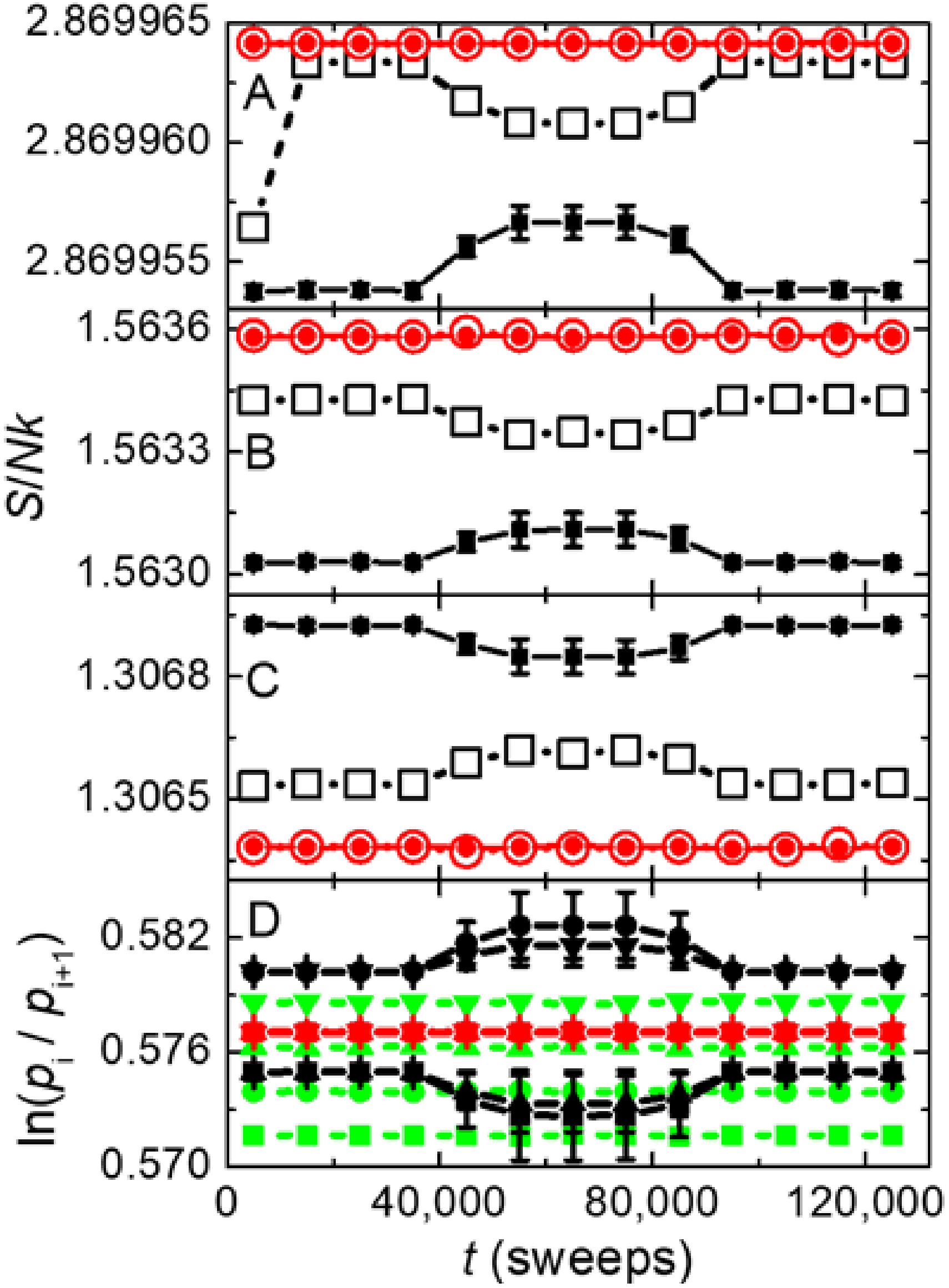

**FIG 2.** Moving averages of $S/Nk$ from the simulations shown in Fig. 1 for spins (C), bath (B), and combined (A). Symbol style identifies dynamics as reversible (black squares) or irreversible (red circles), from initial simulations (open) or from averaging over subsequent simulations (filled), with error bars from the standard error. During the middle third of each simulation the average rate of bond-change attempts is reduced to $1/10^{th}$ the rate of spin-change attempts. (D) Moving averages of the ratio of probabilities between adjacent energy levels in $K$. Symbol shape identifies the levels: $i = 0$ (squares), $i = 1$ (circles), $i = 2$ (up triangles), and $i = 3$ (down triangles). Black and red symbols are from the simulations shown in A-C having reversible and irreversible dynamics, respectively. Green symbols are from irreversible dynamics in a system having $N = 125$.

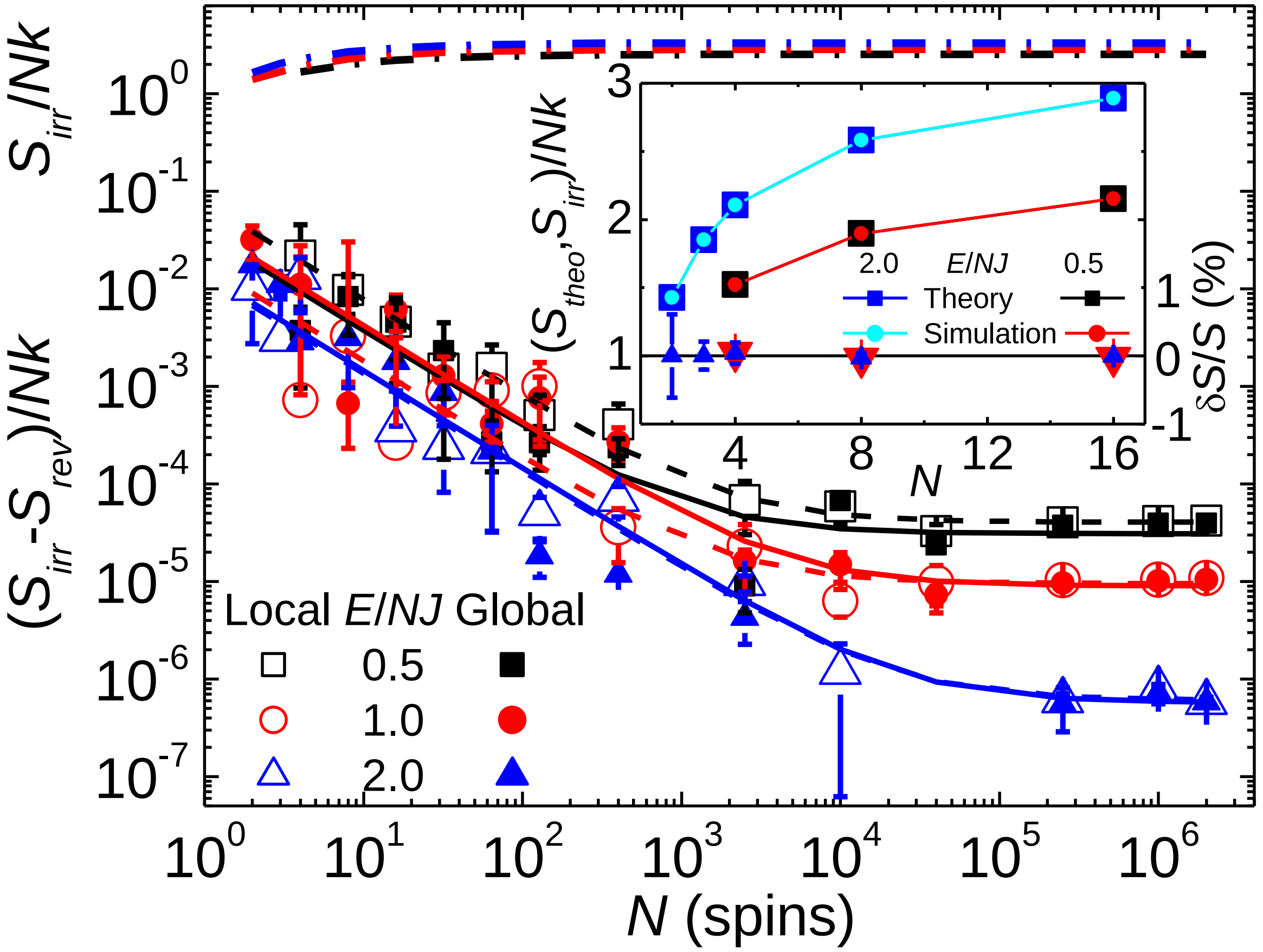

**FIG 3.** Entropies, and their differences, as a function of $N$ on a double-logarithmic scale (main figure) and linear scale (inset). Colors identify the total energy, given in the legends. In the main figure, dash-dotted lines show $S_{irr}/Nk$ from irreversible simulations; symbols show $\Delta S/Nk = (S_{irr} - S_{rev})/Nk$, with the bath for reversible dynamics from local $K$ (open) or global $K$ (filled). Error bars are from the standard error of the mean. Lines show best fits to $\Delta S/Nk = s/N + c$ for local (dashed) and global (solid) $K$. The inset shows behavior from small systems. Symbols give $S_{irr}/Nk$ from simulations (circles) and $S_{theo}/Nk$ from theory (squares). Triangles show the percent difference between theory and simulation, $(S_{irr} - S_{theo})/S_{theo}$, using the right-hand scale.

**Supplemental Material**

We simulate Creutz-like models as follows. Each simulation utilizes a 1D ring of $N$ spins and $N$ sources of $K$, with one source at each spin site. The spins are initialized into a low-energy configuration, with the first half of the spins in the ring oriented up and the second half oriented down. This configuration allows simulations at relatively low energies without the risk of freezing all spins into their fully-aligned state. $K$ is initialized by adding units of energy to randomly-chosen sites until $E$ reaches the set value for the simulation. A single step in the simulation proceeds by first choosing a site, then attempting to change the spin and bond associated with that site. Changes that may occur include inverting the spin and/or making or breaking the bond with a neighboring spin. Such changes occur if-and-only-if energy can be conserved. For example, if making the change increases $U$, there must be sufficient energy at the site to keep $K \geq 0$ after the change, consistent with demons acting as sources of kinetic energy. Alternatively, if the attempted change does not increase $U$ the step always proceeds, with any reduction in $U$ given to $K$. Such steps are repeated $N$ times to yield a single sweep. The sweeps are repeated many times, often yielding ~$10^{11}$ steps per simulation. For post-processing evaluation, a set of several quantities is recorded in a file. Each recorded set includes a moving average of $U/N$, $K/N$, $B/N$, and $p_i$ of the lowest eight levels of $K$. For convenience, file size is limited to a total of $2^{17} = 131{,}072$ sets of data per simulation. For short simulations (or large systems), the data set is recorded after every sweep, whereas for long simulations of smaller systems, each quantity is averaged over 10, 100, 1000, or 10,000 sweeps before recording.

We intermix reversible and irreversible simulations to study both types of dynamics. Specifically, the first simulation is reversible, so that the system returns to its far-from-equilibrium starting state at the end of the simulation. The second simulation continues from the final state of the first simulation, redistributes $K$, then proceeds irreversibly. The third simulation continues from the final state of the second simulation, redistributes $K$, then proceeds using the reversible dynamics of the first simulation. Such simulations are repeated to yield 6-10 simulations. Initial evolutions of reversible and irreversible dynamics come from the 1st and 2nd simulations, respectively. Equilibrium thermal behavior comes from averaging subsequent simulations, e.g. 3, 5, 7, and 9th simulations for reversible dynamics and 4, 6, 8, and 10th simulations for irreversible dynamics.

For irreversible dynamics we utilize the Mersenne Twister pseudo-random number generator [32] that is seeded randomly for effectively random behavior. First, a spin site is chosen randomly from the lattice. Then, the $K$ that will govern the dynamics is chosen randomly. Often, this single source of $K$ governs attempts to change the spin and its bond, in random order. Other simulations utilize two randomly-ordered sources of $K$, one for attempts to change the spin and the other for its bond. Because all such simulations show similar results, we identify the crucial ingredient to be the continually varying choice of which source of $K$ is used for each attempt.

All algorithms used for simulating reversible dynamics involve creating arrays of randomly-ordered (but fixed) sequences of sites, with each site occurring once (and only once) in each array. The sequences are chosen randomly to reduce correlations, but fixed so that the dynamics can be reversed by reversing the sequence. Here we focus on two algorithms. The "local $K$" algorithm utilizes a single array for the site to be updated each step, with the single $K$ at that site governing the attempt to first change the spin and then its bond. The "global $K$" algorithm utilizes a second randomly-ordered array for the spin-change $K$ and a third randomly-ordered array for the bond-change $K$, with alternating spin-first or bond-first attempts. The entropy increases slightly by tripling the number of randomly-ordered arrays, but the increase is small indicating that no reversible algorithm can match the maximum entropy of irreversible dynamics.

---

[32] Matsumoto, M.; Nishimura, T. Mersenne Twister: A 623-dimensionally equidistributed pseudo-random number generator. *ACM Trans. Modeling and Computer Simulation* **8**, 3-30 (1998).